\documentstyle[12pt,twoside,fleqn,espcrc1,epsf]{article}

\newcommand{\eq}[1]{eq.~(\ref{#1})}

\newcommand{\ur}[1]{(\ref{#1})}

\newcommand{\fig}[1]{Fig.~\ref{#1}}

\newcommand{\beq}{\begin{equation}}
\newcommand{\eeq}{\end{equation}}

\newcommand{\bI}{\bar I}

\newcommand{\la}[1]{\label{#1}}

\newcommand{\Tr}{{\rm Tr}\;~}

\newcommand{\DetH}{{\rm Det}_{High}}
\newcommand{\DetL}{{\rm Det}_{Low}}
\newcommand{\II}{$I$'s and $\bar I$'s~}
\newcommand{\Det}{{\rm Det}}
\newcommand{\beqa}{\begin{eqnarray}}
\newcommand{\eeqa}{\end{eqnarray}}
\def\pj{\hspace{-.2cm}}

\def\Dirac#1{#1\hskip-6pt/}
\def\dd{\Dirac\partial}
\def\dD{\nabla\hskip-8pt/}

\newcommand{\AmS}{{\protect\the\textfont2
  A\kern-.1667em\lower.5ex\hbox{M}\kern-.125emS}}

\title{Towards a Theory of Instantons at Non-zero Fermion Density}

\author{Gregory W. Carter
\address{Niels Bohr Institute, Blegdamsvej 17, 2100 Copenhagen \O,
Denmark}
        and Dmitri Diakonov
        \address{NORDITA, Blegdamsvej 17, 2100 Copenhagen \O, Denmark}}

\begin{document}
\maketitle

\begin{abstract}
A formalism to study light fermions at non-zero chemical potential
in the instanton medium is presented. It is then applied to
investigate chiral symmetry restoration and diquark condensation at 
finite baryon density.
\end{abstract}

\section{INTRODUCTION}

The fact that instanton-induced interactions produce attraction
not only in the $\bar q q$ channel (leading to the spontaneous chiral
symmetry breaking) but also in the $qq$ channel (potentially leading
to diquark condensation) has been first realized in ref.
\cite{BeLa}. It becomes especially clear when the number of colours is
$N_c=2$, where instanton-induced interactions possess a global
$SU(4)$ symmetry \cite{DP5} (often referred to as Pauli--G\"ursey
symmetry). For this reason $\bar q q$ and $qq$ condensates
belong in fact to one phase: one condensate can be rotated to
another along the Goldstone valley \cite{DP5}.

The possibility of diquark condensation (in analogy to
superconductivity) at any $N_c$, as induced by instantons, has been
studied in ref. \cite{DFL}. Only a metastable diquark-condensed vacuum
has been found at $N_c>2$ and chemical potential $\mu=0$.
Furthermore, already at $N_c=3$ the scalar diquark appears to be
unbound in the vacuum, indicating that our world is in a sense closer
to the idealized $N_c\rightarrow\infty$ than to the $N_c=2$ world.
Parametrically, the diquark mass is $\sim 1/\bar\rho\sim 1\;{\rm GeV}$ where
$\bar\rho$ is the average instanton size as explained below. It means that a
scalar diquark correlation function should decay with the exponent
corresponding to the `constituent' quark threshold $2M(0)\approx
700\;{\rm MeV}$, which seems to be supported by recent lattice
measurements \cite{HKLW}.  Nevertheless, it has been suggested in
\cite{DFL} that $qq$ condensates could be found as metastable states
in heavy ion collisions and in astrophysics.

Quite recently diquark condensation owing to instantons at $\mu\neq 0$
has been estimated by the Stony Brook and Princeton/MIT groups
\cite{StonyBrook,Princeton/MIT}, see also these proceedings. Below we
present a formalism to study it in a more systematic way.
A theory of light quarks in the instanton vacuum, though not completely
trivial, is rather well developed, and it is straightforward to
generalize it to $\mu\neq 0$. We will avoid some of the unnecessary
approximations made in \cite{StonyBrook,Princeton/MIT}.


\section{QCD INSTANTON VACUUM}

The Euclidean QCD partition function for $N_f$ fermions with a common
chemical potential $\mu$ is
\beq
{\cal Z}=\int\!DA_\mu\exp\left(-\!\int\!\frac{F_{\mu\nu}^{a 2}}{4g^2}
\right)\!\int\!D\psi\:D\psi^\dagger\:\exp\left[\sum_f\!\int\!
\psi_f^\dagger(i\dD+im_f-i\mu\gamma_4)\psi_f\right].
\la{Z1}\eeq
It is understood that the theory is regularized in some way at large
momenta, for example, by the Pauli--Villars method. It is also
convenient to normalize the partition function to the free one. The
integral over fermion fields can be thus understood as
\beq
\!\int\!D\psi\:D\psi^\dagger\:\exp\left[...\right]
\equiv \Det
=\prod_f\frac{\det(i\dD+im_f-i\mu\gamma_4)\det(i\dd+iM)}
{\det(i\dD+iM)\det(i\dd+im_f)}
\la{Z2}\eeq
where $M$ is the Pauli--Villars regulator mass playing the role of the
ultraviolet cutoff.

The main hypothesis of the instanton vacuum is that the partition
function \ur{Z1} is mainly saturated by the ensemble of instantons
and antiinstantons (\II for short), plus quantum fluctuations about
them. Since the ensemble of \II is not, strictly speaking, a solution
of the Yang--Mills equation of motion (it is only in the infinitely dilute
limit), one has to do something more clever than the saddle-point
integration over $A_\mu$. The most straightforward procedure is
to use the Feynman variational principle in order to get the lower
bound for the partition function \ur{Z1} \cite{DP1}. Writing the
general Yang--Mills field as a superposition of $N_+$ $I$'s and
$N_-$ $\bar I$'s plus a presumably weak quantum field $B_\mu$,
\beq
A_\mu=\sum A_\mu^I(\xi)+\sum A_\mu^{\bar I}(\xi)+B_\mu
\la{genA}\eeq
where $\xi$ is a set of $4N_c$ collective coordinates for each
instanton (namely center, size and colour orientation), one can show
that the partition function \ur{Z1} can be approximated by
\beq
{\cal Z}\geq\sum\frac{1}{N_+!N_-!}\int\!d\xi\:J(\xi)\:
\exp\left[-U_{int}(\xi)\right]\cdot \Det.
\la{Z3}\eeq
Here $J$ is a Jacobian which is due to passing to the
collective coordinates and $U_{int}$ is the instanton interaction.
Actually, the separation of the dependence on the collective
coordinates $\xi$ into $J$ and $U_{int}$ is ansatz-dependent since
the precise meaning of the collective coordinates depends on the
functional form of the field. However, their combination should be
ansatz-independent, at least if one calculates both quantities with
infinite precision. We shall assume the simplest sum ansatz \ur{genA}.

It has been shown in ref. \cite{DP1} (see also a more recent paper
\cite{DPW}) that the partition function \ur{Z3} (without the fermion
determinant) leads to the stabilization of the grand canonical
instanton ensemble. The basic characteristics of the medium is the
average instanton size, $\bar\rho$, and the average separation between
near neighbours, $\bar R$, or the instanton density at equilibrium,
$N/V=1/(\bar R)^4$. Since the transmutation of dimensions takes place
in this approach, both quantities are expressed through the only
dimensional parameter one has in the pure glue theory, namely
$\Lambda_{QCD}$. The 2-loop calculations performed in \cite{DP1,DPW}
give
\beq
\bar\rho = \sqrt{\overline{\rho^2}} \approx
\frac{0.48}{\Lambda_{\overline{MS}}}, \;\;\;\;\;
\bar R \approx \frac{1.35}{\Lambda_{\overline{MS}}}.
\la{rhoR}\eeq
Taking $\Lambda_{\overline{MS}}=280\;{\rm MeV}$ one finds
$\bar\rho\approx 0.35\;{\rm fm},\;\;\bar R \approx 0.95\:{\rm fm},
\;\;\bar\rho/\bar R\approx 1/3$. This small ratio has been previously
suggested on phenomenological grounds by Shuryak \cite{Sh1}.
The smallness of the $\bar\rho/\bar R$ ratio implies that
the packing fraction of instantons in the vacuum, i.e. the
fraction of the 4d volume occupied by the balls of radius
$\bar\rho$, is quite small:
\beq
f=\frac{\pi^2}{2}\bar\rho^4\frac{N}{V}
=\frac{\pi^2}{2}\frac{\bar\rho^4}{\bar R^4}\sim\frac{1}{10}.
\la{pack}\eeq

\subsection{Instanton Packing Fraction as Algebraic Parameter}

After the transmutation of dimensions takes place all physical
quantities are proportional to $\Lambda_{QCD}$ in appropriate powers
with numerical coefficients of the order of unity, as it should be,
generally speaking, in a strong interaction theory. Therefore, if
one gets a numerically small (or large) dimensionless quantity it must
be due to some exceptional reason. In our case the reason
why one gets a numerically small packing fraction \ur{pack} can be
traced back to the `accidentally' large value 11/3 of the
coefficient in the asymptotic freedom law. The same analysis applied
to the 2d $CP^N$ model also possessing instantons shows that
there are no such numerical tricks there: instantons of that
model seem to be strongly overlapping, and there appears no
small packing fraction parameter.

Though $\bar\rho/\bar R\approx 1/3$ is but numerically small, one
can treat it as a formal algebraic parameter, and develop a
perturbation theory in it. In principle, one can change the parameter
$\bar\rho/\bar R$ by taking more quark flavours, or taking nonzero
temperature or including Higgs field which cuts off integrals over the
instanton sizes.

\subsection{Separating High and Low Fermion Eigenmodes}

We now switch on $N_f$ flavours of light quarks represented by
the fermion determinant \ur{Z2} in the backround field of the instanton
ensemble. The crucial feature of fermions in the one-instanton
background is that there is an exact zero mode \cite{tH}.
In the infinitely dilute limit of instantons that would lead to the
nullification of the fermion determinant. Though the packing fraction
of instantons is small it is, however, nonzero, and one can build
a systematic theory of light quarks in the instanton vacuum by treating
$\bar\rho/\bar R$ as a formal small parameter \cite{DP2,DP3,DP4}.

An important step is to separate contributions of high- and
low-frequency parts to the fermion determinant. The role of the
high-frequency part is to renormalize somewhat the one-instanton
weight or fugacity. The low-frequency part is physically much more
interesting: it is responsible for chiral symmetry breaking in
the vacuum and its restoration at high densities and/or temperatures.

The separation is made in respect to an auxiliary mass $M_1$ lying
inside a parametrically wide interval, $1/\bar R\ll M_1\ll 1/\bar\rho$.
One rewrites the fermion determinant \ur{Z2} as a product,
\beq
\Det =
\frac{\det(i\dD+im_f-i\mu\gamma_4)\det(i\dd+iM)}
{\det(i\dD+iM)\det(i\dd+im_f)}
=\DetH\cdot \DetL\,,
\la{prod}\eeq
where
\beqa
\DetH=
\frac{\det(i\dD+iM_1)\det(i\dd+iM)}
{\det(i\dD+iM)\det(i\dd+iM_1)},\;\;\;
\DetL=
\frac{\det(i\dD+im_f-i\mu\gamma_4)\det(i\dd+iM_1)}
{\det(i\dD+iM_1)\det(i\dd+im_f)}.
\nonumber\eeqa
These ratios are arranged so that $\DetH$ gets a contribution from
fermion modes with Dirac eigenvalues ranging from $M_1$ to the
Pauli--Villars mass $M$, while $\DetL$ is saturated by eigenvalues
less than $M_1$. The product of the two determinants is, by
construction, independent of the separation scale $M_1$. However,
we are going to treat both of them approximately. The fact that
the actual dependence of the product on $M_1$ turns out to be
extremely feeble \cite{DP2} in the wide range of $M_1$
serves as a check of the approximations.

In considering $\DetH$, one assumes that $M_1$ is large enough so
that $\DetH$ can be factorized into a product of determinants
computed in the background of individual instantons. Corrections
to the factorization are of the order of $(M_1\bar R)^{-2}\ll 1$.

In considering $\DetL$, one assumes that $M_1$ is small enough
so that only the (diagonalized) would-be zero modes of individual
instantons can be taken into account. Corrections from nonzero
modes of individual instantons are of the order of
$(M_1\bar\rho)^2 \ll 1$.

Taking nonzero chemical potential $\mu$ does not seriously change the
above logic. At $\mu\bar\rho\ll 1$ (which will be our domain of
interest since all important phenomena happen inside this range) the
$\mu$ term should be included into the `low' part as in \eq{prod}.

\subsection{Three Ways to See Chiral Symmetry Breaking}

In refs.\cite{DP2,DP3,DP4} three seemingly different but actually
equivalent ways of computing $\DetL$ have been developed, thus leading
to several languages in which chiral symmetry breaking by instantons
can be understood and mathematically described.

The first method is based upon diagonalizing the overlaps of the
would-be zero
fermion modes of individual instantons \cite{DP2,DP3}.  One can write
\beq
\DetL=\exp\left\{\frac{1}{2}\int\!d\lambda\:\nu(\lambda)
\ln\frac{\lambda^2+m^2}{\lambda^2+M_1^2}\right\}
\la{spectrdens}\eeq
where $\nu(\lambda)$ is the spectral density of the Dirac operator
in the field of the instanton ensemble. It is obtained from
diagonalizing a matrix made of the overlap integrals of the would-be
zero modes, and then averaging over the ensemble. The nonzero
$\nu(0)$ signals chiral symmetry breaking: it is actually due to
the delocalization of the zero modes owing to the ``hopping''
of quarks from one instanton to another.

The second method is to find the quark propagator in the instanton
ensemble. In the zero-mode approximation (justified when the
packing fraction \ur{pack} is small) one can write a {\it closed}
equation for the quark propagator averaged over the instanton
ensemble \cite{DP3,Pob}. The propagator appears to be that
of a massive fermion with a momentum-dependent mass $M(p)$ related to
the Fourier transform of the zero mode. Using the main characterictics
of the instanton vacuum \ur{rhoR} one gets $M(0)\approx 350\;{\rm MeV}$
which is what one expects for the so-called constituent quark mass.
 Parametrically, $M(0)\sim\sqrt{N/VN_c}\bar\rho$, meaning
that the constituent quark mass is small in the packing fraction
of instantons.

Integrating the quark propagator over the 4-momentum one gets the
chiral condensate $\langle \bar\psi \psi\rangle\approx -(250\;{\rm
MeV})^3$ being also close to its phenomenological value.

The $\DetL$ can be expressed through the propagator $S(x,y)$ as
\beq
\DetL=\exp\int_m^{M_1}\!i\:dm^\prime\:\Tr\!\left[S(x,x,m^\prime)-
S_0(x,x,m^\prime)\right]
\la{detprop}\eeq
where $S_0$ is the free propagator. Comparing the integrand in
\eq{detprop} with that of \eq{spectrdens} one finds the
Wigner semicircle spectral density of the Dirac operator averaged over
the instanton ensemble \cite{DP3},  
which can also be obtained directly from the first method \cite{Sim}.

Both methods can be directly generalized to nonzero chemical potential,
since at $\mu\neq 0$ there also exists an exact normalizable zero
fermion mode \cite{Abr,DeCarv}. However, in this paper we use the third
method suggested in \cite{DP4} which is also easily translated to
$\mu\neq 0$.

\section{EFFECTIVE FERMION ACTION AT $\mu\neq 0$}

Rather than calculating fermionic quantities in the background of
instantons and then averaging over these configurations,
one can perform the averaging procedure first. This generates
instanton-induced interactions in the form of a vertex involving
$2N_f$ quarks. It is a direct consequence of the appearance of the
quark zero modes, one for each flavour, and the resulting interactions
are naturally of the 't Hooft form \cite{tH}.

The guiding principle is to write a fermion partition function
reproducing the propagator at low momenta as well as the fermion
determinant. This derivation has been developed in detail in previous
publications \cite{DP4} and its extension to finite chemical potential
is straightforward. It should be stressed that correlations between
instantons induced by fermions are inherent in this approach; as to
correlations induced by gluons, they are effectively taken care of by
the use of the variational principle \cite{DP1} resulting in the
effective size distribution. For simplicity we freeze all the sizes at
the average value $\bar\rho$, but average explicitly over randomly
positioned and oriented instantons.

The partition function which produces the necessary propagator is of
the form
\beq {\cal Z}
=\DetL=\int\!D\psi\:D\psi^\dagger\:\exp\left[\sum_f\!\int\!
\psi_f^\dagger(i\dd-i\mu\gamma_4)\psi_f\right]
\left(\frac{Y^+_{N_f}}{V} \right)^{N_+} \left( \frac{Y^-_{N_f}}{V}
\right)^{N_-}.
\la{Z4}\eeq
The pre-exponential factors contain the instanton-induced interactions
between fermions, and for $N_f$ flavours are the specific nonlocal
$2N_f$-fermion vertices
\beq
Y^{\pm}_{N_f}\! = \!(-)^{N_f}\!\!
\int\!\!d^4z\:dU\! \prod^{N_f}\!\int\!d^4x\:d^4y\:
\psi^\dagger_{L,R}(x)(i\partial-i\mu)^{\mp}\Phi_{\bar I,I}(x-z)
\tilde\Phi_{\bar I,I}(y-z)(i\partial-i\mu)^{\pm}\psi_{L,R}(y) .
\eeq
Here we use the notation $x^{\pm}=x^\mu\sigma_\mu^{\pm}$, where the $2
\times 2$ matricies $\sigma_\mu^{\pm} = (\pm i \vec \sigma,1)$ decompose
the Dirac matricies into chiral components,
and it is understood that $\mu$ written as a four-vector is
$\mu_\alpha = (\vec 0,\mu)$.  Note that the zero mode in the field of
an (anti) instanton couples to that of a (right-) left-handed quark.
By $\tilde \Phi_{I(\bI)}$ we denote the conjugate zero mode. As noticed
in \cite{Abr}, $\tilde \Phi_I(x,\mu)=\Phi_I^\dagger(x,-\mu)$, where
the dagger means hermitean conjugate.

Fermion operators in the pre-exponent are not
convenient; these operators can be raised into the exponent
with the help of a supplementary integration over a pair of Lagrange
multipliers, denoted $\lambda_{\pm}$:
\begin{eqnarray}
{\cal Z} &\pj=&\pj \int\!d\lambda_+d\lambda_-\int
\!D\psi\:D\psi^\dagger\:\exp\Bigg\{\int\!d^4x\:\psi^\dagger(i\dd-i\mu\gamma_4)
\psi + \lambda_+ Y^+_{N_f} + \lambda_- Y^-_{N_f}\nonumber\\
&& \quad\quad\quad\quad + N_+\left(\ln\frac{N_+}{\lambda_+ V} -1 \right) +
 N_-\left(\ln\frac{N_-}{\lambda_- V} -1 \right) \Bigg\}.
\label{Z5}
\end{eqnarray}
 Indeed, integrating over $\lambda_\pm$ by the saddle-point
method one recovers \eq{Z4} (the saddle-point integration becomes exact
in the thermodynamic limit $N_\pm\rightarrow\infty$).  Through this
procedure we obtain a purely exponential integrand which is 
the required effective action. It is very important that the overall strength
of the $2N_f$-fermion interaction, whose role is played by
$\lambda_\pm$, is not fixed once and forever: actually its value
is found from minimizing the free energy {\em after} integration over
fermions is performed. Therefore the strength of the interaction
depends itself on the phase the fermion system assumes.
As will be shown below, this is the mechanism by which chiral symmetry is
restored at large chemical potential.

 Although one can investigate the phenomena of chiral symmetry
breaking and its restoration for any number of flavours, we shall
restrict ourselves to the case $N_f = 2$ corresponding to a system of
chiral up and down quarks.  Furthermore, we will assume the CP
invariant case of $\theta=0$, which requires $N_+ = N_- = N/2$ and
hence $\lambda_+=\lambda_-=\lambda$.  For practical applications it is
favourable to use the Fourier-transformed expressions of the quark zero
modes, which are written explicitly in the Appendix.  These complex
functions of the four-momenta and chemical potential determine the
(matrix) form factors attached to each fermion leg of the vertex:
\beq
{\cal F}(p,\mu) = (p+i\mu)^- \varphi(p,\mu)^+\, , \quad\;\;\;
{\cal F}^\dagger(p,-\mu) = \varphi^*(p,-\mu)^-(p+i\mu)^+\,.
\la{calF}\eeq
With such definitions, the interaction terms may be written in momentum space,
\beqa
\lambda_+Y^+ &\pj=&\pj \lambda_+
\int\!\frac{d^4p_1d^4p_2d^4k_1d^4k_2}{(2\pi)^{16}}
(2\pi)^4\delta^4(p_1+p_2-k_1-k_2) \nonumber\\
&&\cdot \int\!dU\:
\psi^\dagger_{L1\alpha_1 i_1}(p_1)
{\cal F}(p_1,\mu)_{k_1}^{i_1}\epsilon^{k_1l_1} U_{l_1}^{\alpha_1}
 U_{\beta_1}^{\dagger o_1} \epsilon_{n_1o_1}
 {\cal F}^\dagger(k_1,-\mu)_{p_1}^{n_1}\psi_L^{1\beta_1p_1}(k_1)
\nonumber \\
&&\cdot \psi^\dagger_{L2\alpha_2 i_2}(p_2)
{\cal F}(p_2,\mu)_{k_2}^{i_2}\epsilon^{k_2l_2} U_{l_2}^{\alpha_2}
 U_{\beta_2}^{\dagger o_2} \epsilon_{n_2o_2}
 {\cal F}^\dagger(k_2,-\mu)_{p_2}^{n_2}\psi_L^{2\beta_2p_2}(k_2)\,,
\la{vertex}\eeqa
with a similar form for $\lambda_-Y^-$ which carries right-handed
quarks. The first indices on the fermion operators refer to flavour (1
or 2 explicitly), the Greek to colour ($1\dots N_c$), and the last
denote spin ($1,2$).  This formulation of the effective
interaction retains the full $p$ and $\mu$ dependence of the zero modes,
as opposed to the approximated treatments in other recent works
\cite{StonyBrook,Princeton/MIT}.

\section{COMPETITION BETWEEN $\bar q q$ AND $qq$ CHANNELS}

Since the instanton-induced interactions \ur{vertex} support both
$\bar q q $ and $q q $ condensation, it is necessary to consider the two
competing channels simultaneously. This means that one must calculate
both the normal ($G$) and anomalous ($F$) quark Green functions. A
colour/flavour/spin ansatz compatible with the possibility of chiral
and colour symmetry breaking is
\beqa
\langle\psi^{f\alpha i}(p)\psi^\dagger_{g\beta j}(p)\rangle &\pj=&\pj
\delta^f_g \delta^\alpha_\beta S_1(p)^i_j \quad
{\rm for}\:\alpha,\beta=1,2 \,,
\nonumber
\\
\langle\psi^{f\alpha i}(p)\psi^\dagger_{g\beta j}(p)\rangle &\pj=&\pj
\delta^f_g \delta^\alpha_\beta S_2(p)^i_j \quad
{\rm for}\:\alpha,\beta>2 \,,
\nonumber
\\
\langle\psi_L^{f\alpha i}(p)\psi_L^{g\beta j}(-p)\rangle
&\pj=&\pj
\langle\psi_R^{f\alpha i}(p)\psi_R^{g\beta j}(-p)\rangle
= \epsilon^{fg}\epsilon^{\alpha\beta[\gamma]}\epsilon^{ij} F(p)\,,
\eeqa
where $[\gamma]$ refers to some generalized direction(s) in colour space,
and it is this set of $N_c-2$ indices which signals the breaking of
colour symmetry.  In the particular case of $N_c=3$, where the colour
symmetry is broken as $SU(3)\rightarrow SU(2)\times U(1)$, we will by
convention take $[\gamma]=3$; for $N_c=4$ one can take
$[\gamma]=34$  and so forth. In the event of colour symmetry
breaking, the standard propagators (and ensuing condensates) will lose
their colour degeneracy and the separation of $S(p)$ into $S_1(p)$ and
$S_2(p)$ becomes necessary; otherwise the Schwinger-Dyson equations do
not close.

Written in the chiral $L,R$ basis, the $4\times 4$ propagators
$S_{1,2}(p)$ are of the form:
\beq
S(p) = \left[\begin{array}{cc}G(p) {\bf 1} & Z(p){\bf S}_0(p)^+ \\
Z(p){\bf S}_0(p)^- & G(p) {\bf 1} \end{array}\right] \,,\quad
{\bf S}_0(p)^{\pm} = \left[(p+i\mu)^\pm\right]^{-1}\;.
\eeq
Here the off-diagonal, bare propagator is modified by the scalar functions
$Z_{1,2}(p)$, and is augmented on the diagonal by the scalar
$G_{1,2}(p)$ which if nonzero break chiral symmetry.

\begin{figure}[tb]
\setlength\epsfxsize{14cm}
\centerline{\epsfbox{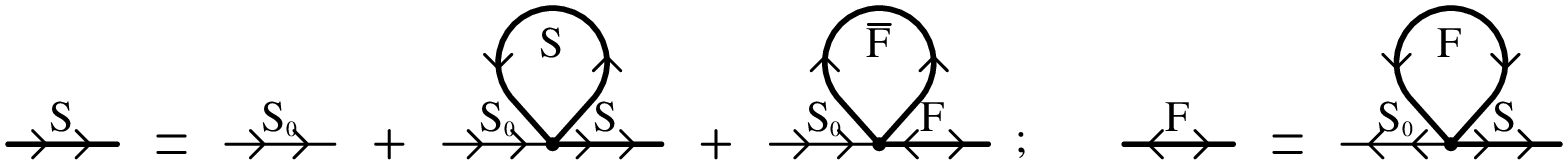}}
\vskip-.8cm
\caption{Gorkov equations to first order in $\lambda$.}
\label{gorkfig}
\end{figure}

Using the instanton-induced interaction \ur{vertex} one can build a
systematic expansion for the $F,G$ Green functions in the $1/N_c$
and $\bar\rho/\bar R$ parameters. In the leading order in both
parameters we restrict ourselves to the one-loop approximation
shown in \fig{gorkfig}. It corresponds to a set of self-consistent
Schwinger-Dyson equations (called Gorkov equations in the case of
superconductivity with $F\neq 0$).
An important $\mu$-dependence enters through the form factors in
\eq{vertex}.
With $\bar F(p)=F^*(p)$ these diagrams lead to the set of five
algebraic equations for the scalar Green functions $Z_{1,2},\,G_{1,2}$
and $F$:
\beqa Z_1(p) &\pj=&\pj 1 - G_1(p) A(p,\mu)\bar{g}_1 -
F(p)B(p,\mu)\bar{f}\nonumber\\ Z_2(p) &\pj=&\pj 1 - G_2(p)
A(p,\mu)\bar{g}_2 \nonumber\\ G_1(p) &\pj=&\pj Z_1(p)
\varphi_\alpha(p,\mu)\varphi_\alpha(p,\mu)\bar{g}_1 \nonumber \\ G_2(p)
&\pj=&\pj Z_2(p) \varphi_\alpha(p,\mu)\varphi_\alpha(p,\mu)\bar{g}_2
\nonumber\\
F(p) &\pj=&\pj Z(-p) \varphi_\alpha(p,\mu)\varphi_\alpha(-p,\mu)\bar{f}\,.
\la{five}\eeqa
The numerical factors $\bar{g}_1$, $\bar{g}_2$, and $\bar{f}$ will be
defined below and the functions
\beqa
A(p,\mu) &\pj=&\pj (p+i\mu)_\alpha(p+i\mu)_\alpha \varphi_\beta(p,\mu)
\varphi_\beta(p,\mu) \,, \nonumber\\
B(p,\mu) &\pj=&\pj (p^2+\mu^2)\varphi_\beta(p,\mu)\varphi_\beta(-p,\mu)
+ (p+i\mu)_\alpha\varphi_\alpha(p,\mu)(p-i\mu)_\beta\varphi_\beta(-p,\mu)
\nonumber\\
&&\, - (p+i\mu)_\alpha\varphi_\alpha(-p,\mu)(p-i\mu)_\beta\varphi_\beta(p,\mu)
\,, \label{GFs}
\eeqa
are the form factors which arise from the zero modes (see the
Appendix). At $\mu =0$ we have $A(p,0)=B(p,0)$,
but for any finite $\mu$ the direction of the momentum flow
through each vertex leg is critical.

The condensates $g_1$, $g_2$, and $f$ are the closed loops
contributing to the quark self-energy.  They are found by integrating the
appropriate Green function, modified by the vertex form factors, over
an independent momentum:
\beqa
g_{1,2} &\pj=&\pj \frac{\lambda}{N_c^2-1} \int\!\frac{d^4k}{(2\pi)^4}\:
A(k,\mu)G_{1,2}(k)
\, , \quad
f = \frac{\lambda}{N_c^2-1} \int\!\frac{d^4k}{(2\pi)^4}\: B(k,\mu)F(k) \,.
\label{condefs}
\eeqa
These appear in the particular combinations
\beqa
\bar{g}_1 &\pj=&\pj \left(5-\frac{4}{N_c}\right)g_1+
\left(2 N_c-5+\frac{2}{N_c}\right)g_2
\, , \quad
\bar{g}_2 = 2\left(2-\frac{1}{N_c}\right)g_1+2(N_c-2)g_2\,,\nonumber\\
\bar{f} &\pj=&\pj 2\left(1+\frac{1}{N_c}\right)f\,.
\eeqa
Although the integrands above are complex, all imaginary parts are
odd in $p_4$ and thus vanish under integration.
As the $\bar g_{1,2}$ are measures of chiral symmetry breaking, these
act as an effective mass modifying the standard quark propagation.  On the
other hand the diquark loop $\bar f$ plays a different role,
that of twice the single-quark energy gap (conventionally denoted
$\Delta$) formed around the Fermi surface. The Fermi
momentum, in the absence of chiral symmetry breaking, will remain at
$p_f=\mu$ regardless of the magnitude of $\bar f$.

After determining the five scalar functions through solving eqs. \ur{five}
and inserting these solutions into eqs. \ur{condefs} we find the coupled
equations for the condensates themselves:
\beqa
g_1 &\pj=&\pj \frac{\lambda \bar{g}_1}{N_c^2-1}\int\!\frac{d^4k}{(2\pi)^4}\:
\frac{\alpha(1-\beta\bar{f}^2+\alpha^*\bar{g}_1^2)}
{(1+\alpha\bar{g}_1^2)(1+\alpha^*\bar{g}_1^2)-\beta^2\bar{f}^4}\,
\,,\quad
g_2 = \frac{\lambda\bar{g}_2}{N_c^2-1}\int\!\frac{d^4k}{(2\pi)^4}\:
\frac{\alpha}{1+\alpha\bar{g}_2^2}\,,\nonumber\\
f &\pj=&\pj \frac{\lambda\bar{f}}{N_c^2-1}\int\!\frac{d^4k}{(2\pi)^4}\:
\frac{\beta(1-\beta\bar{f}^2+\alpha\bar{g}_1^2)}
{(1+\alpha\bar{g}_1^2)(1+\alpha^*\bar{g}_1^2)-\beta^2\bar{f}^4}\,,
\label{CONDS}
\eeqa
where yet another pair of functions has been introduced,
\beq
\alpha(p,\mu) = A(p,\mu)\varphi_\alpha(p,\mu)\varphi_\alpha(p,\mu)\,,\quad
\beta(p,\mu) = B(p,\mu)\varphi_\alpha(p,\mu)\varphi_\alpha(-p,\mu)\,.
\eeq
Note that while $\beta$ is real, the funtion $\alpha$ is complex.  As usual
for gap equations, there is the possibility of a solution where some (or
even all) of the condensates vanish.

\begin{figure}[bt]
\setlength\epsfxsize{8cm}
\centerline{\epsfbox{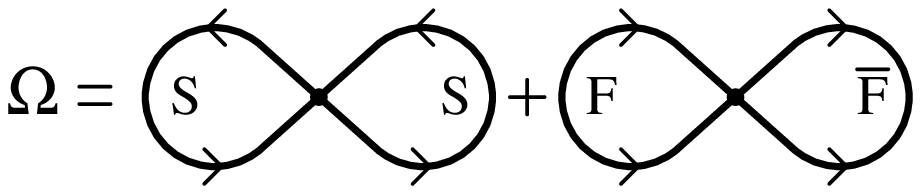}}
\vskip-1.2cm
\caption{Contributing diagrams to $\Omega$ at order $\lambda^1$.}
\label{eightfig}
\end{figure}

The magnitudes of these condensates are not yet determined, as this
requires fixing the coupling constant $\lambda$.  This is done through
minimizing the partition function \ur{Z5} and leads to
\beq
\frac{N}{V} = \lambda\langle Y^+ + Y^- \rangle \,.
\eeq
On the left-hand side is the instanton density, which we will here take
to be fixed at its vacuum value, although in principle will have some
correction due to the finite quark density.  Evaluating the right-hand
side requires calculating the one-vertex contributions to the free
energy, which in this case includes two-loop, figure-eight type diagrams
formed by joining the four fermion legs into two pairs as
shown in \fig{eightfig}.  This can be
written concisely in terms of the condensates:
\beq
\frac{N}{V} = \lambda\langle Y^+ + Y^-\rangle = \frac{4(N_c^2-1)}{\lambda}
\left[ 2 g_1 \bar{g}_1 + (N_c-2) g_2\bar{g}_2 + 2 f\bar{f} \right]\,.
\label{gapeqn}
\eeq
This equation for $\lambda$ and the definitions of the condensates
\ur{CONDS} comprise a system of equations which can be solved for all
quantities.

Once this has been achieved the chiral condensate may be calculated,
which is a closed trace over the quark propagator.  This is
distinct from the condensates $\bar{g}_{1,2}$, however it does contain
the same properties under $p_4$ reflection which guarantee it is real.
It is most simply written
\beq -\langle\bar{\psi}\psi\rangle_{Mink} =
i\langle\psi^\dagger\psi \rangle_{Eucl} =
i\int\!\frac{d^4p}{(2\pi)^4}\,\Tr{S(p)} =
4\int\!\frac{d^4p}{(2\pi)^4}\,\left[2 G_1(p) + (N_c-2) G_2(p)\right] .
\eeq

As a consequence of having two different modes of quark condensation,
one obtains multiple solutions for the characteristics of the quark
medium at any fixed chemical potential.  Specifically, there will be a
competition between the following phases:

(0)  No condensation of any type; $g_1=g_2=f=0$.

(1)  Chiral symmetry breaking without diquark condensation.  This, the
standard vacuum scenario, has the signature $g_1=g_2\ne 0$, $f=0$.

(2)  Diquark condensation without chiral breaking, or $f\ne 0$,
$g_1=g_2=0$.

(3)  A mixed phase wherein both symmetries are broken, and $g_1$, $g_2$,
and $f$ are all distinct and finite.

Phase (0) requires that $\lambda$ vanish, and such a solution was
never found in our calculations.
To resolve between the remaining, symmetry-breaking phases, the free energy is
minimized. Consistent with the
evaluation of the Green functions, it is calculated to first order in
$\lambda$. Repeating the calculation of figure-eight diagrams and
recalling the explicit dependence on $\lambda$ in \eq{Z5} we obtain the
free energy
\beq
\frac{\Omega}{V_3} = -\frac{1}{\beta V_3}\ln Z
= \frac{\Omega_0}{V_3}-\frac{N}{V}\ln\left(\frac{N}{\lambda V}\right)
+ \frac{N}{V}
-\frac{4(N_c^2-1)}{\lambda}
\left[ 2 g_1 \bar{g}_1 + (N_c-2) g_2\bar{g}_2 + 2 f\bar{f} \right].
\eeq
Here $V_3$ is the three-volume while $V$ is the Euclidean four-volume, and
$\Omega_0$ is the free energy for a gas of free quarks.
The last two terms are precisely the quantity which must vanish under the
saddle-point determination of $\lambda$, and thus we have
\beq
\frac{\Omega}{V_3} = \frac{\Omega_0}{V_3}+\frac{N}{V}
\ln\left(\frac{\lambda}{N/V}\right)\,.
\eeq
Thus the phase which features the {\em lowest} coupling $\lambda$
is the thermodynamically favoured.

Numerically, no solutions were found which favour phase (3), the mixed
case. Thus we concentrate on the distinction between phases
(1) and (2), the former being marked by nonzero $g=g_1=g_2$,
the latter by nonzero $f$. Each case leads to its own value of
$\lambda$ through solving \eq{gapeqn} for any given $\mu$.
At certain value of $\mu=\mu_c$ the solutions for $\lambda$
corresponding to the two phases cross; this is the point where
the phase transition occurs. This point is defined by the condition
\beq \frac{f(\mu_c)}{g(\mu_c)} = \sqrt{\frac{N_c(N_c-1)}{2}}\,.
\label{CR} \eeq
When the ratio between the condensates is less than the constant on the
right, phase (2) is favoured; otherwise it is chiral symmetry that is
spontaneously broken in phase (1).  Calculations were carried out for
various $N_c$, taking the values for $N/V$ and $\bar\rho$ specified in
Section 2.

\subsection{The Case of $N_c$ = 2}

In this case it is obvious that colour symmetry is not broken by diquark
formation, which here correspond to colour-singlet `baryons', and hence
there is only one possible chiral condensate $g_1=g$.
Not so obvious is that at $\mu =0$ the colour and flavour
$SU(2)$ groups are arranged into the higher $SU(4)$ symmetry.
The instanton vacuum accounts for this symmetry \cite{DP5}, and in 
the context of the analysis here this corresponds to
$f^2 + g^2$ being the only discernable quantity in the gap
equations \ur{CONDS}. Numerically, we find $\sqrt{f^2+g^2}= 147$ MeV.

At finite $\mu$, however, the $SU(4)$ symmetry is explicity broken and
one finds $f<g$. Since in this case the
critical ratio of \eq{CR} is unity, we conclude that for any finite
density the $N_c=2$ world prefers diquark condensation to chiral
symmetry breaking.  This finding is in agreement with the reasoning of
ref. \cite{StonyBrook} and the lattice results of ref. \cite{Nc2Lattice}.

\begin{figure}[tb]
\setlength\epsfxsize{13cm}
\centerline{\epsfbox{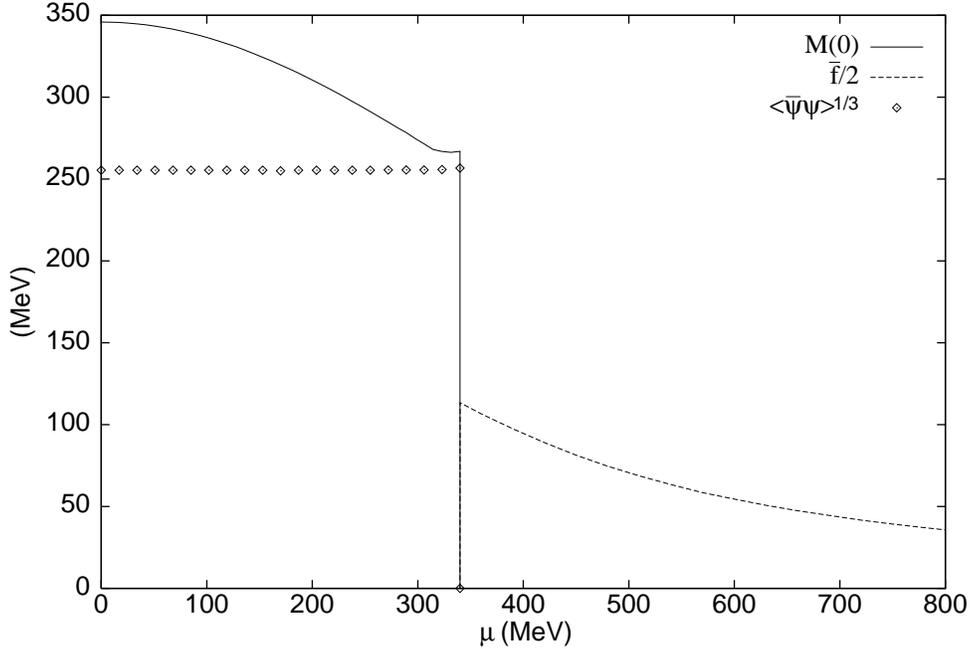}}
\vskip-1.2cm
\caption{Condensates for $N_c=3$ as a function of $\mu$.
Shown are the effective quark mass $M(0)$, the quark condensate
$\langle\bar{\psi}\psi\rangle^{1/3}$, and the diquark energy gap
per quark $\bar{f}/2$.} \label{plot} \end{figure}

\subsection{The Case of $N_c$ = 3}

For three colours, the critical ratio of the condensates
\ur{CR} is $\sqrt{3}$. At $\mu =0$ we find
$f/g=165\,{\rm MeV}/65\,{\rm MeV} > \sqrt{3}$. It means that at
low values of $\mu$ the coupling constant $\lambda$ is smaller in
the chiral broken phase, hence this phase is energetically preferred.
Furthermore, the effective quark mass is the reasonable $M(0) = \bar{g} 
= 346$ MeV. We thus recover the standard chiral symmetry breaking
at low baryon density.

With increasing chemical potential both $f$ and $g$ condensates in their
respective phases are reduced, however as $\mu$ surpasses the effective
quark mass $M(0)$ around 300 MeV, the chiral-breaking $g$ stabilizes
while the colour-breaking $f$ continues to decrease in phase (2).  At
$\mu\simeq 340$ MeV condition \ur{CR} is reached, and therefore for this
and greater chemical potential chiral symmetry is restored and the
`colour superconducting' phase is realized.  At the transition point,
the quark gap is $\bar{f}/2 = 115$ MeV and slowly decreasing with
rising density.  Numerical results for the characteristic quantities of
each phase are plotted in \fig{plot}.  It is noteworthy that the chiral
condensate, $\langle\bar{\psi}\psi\rangle$, is distinct from the
effective quark mass, $\bar{g}$, in that while the latter decreases
with $\mu$ the former remains practically at its vacuum value. For
$\mu > \mu_c$ they both vanish.

\section{CONCLUSIONS}

We have formulated the effective low-energy fermion action
induced by instantons at nonzero chemical potential.  In the resulting
fermion vertex we have retained the full dependence on both momentum
and chemical potential, which arise from the would-be zero modes.  In
this respect we differ from other studies such as the random matrix
model, NJL models, and alternative instanton approaches.

The effective action leads to a competition between two phases, one of
chiral symmetry breaking and another characterized by diquark
condensation.  It was studied by solving a coupled system of gap
equations to first order in the instanton density.  For two massless
flavours and three colours, it was found that the usual broken chiral
symmetry is restored through a first order phase transition,
replaced by colour breaking due to the formation of a diquark
condensate.  This occurs at a critical chemical potential of $\simeq
340$ MeV.

\section*{APPENDIX. Fourier Transforms of Fermion Zero Modes}

The use of the exact fermion zero modes in the momentum space
tremendously simplifies all calculations. The starting point is
the exact fermion zero mode in the field of one (anti)instanton
in $x$ space \cite{Abr,DeCarv}, which we cite for arbitrary
instanton position $z$, size $\rho$ and orientation given by
rectangular $N_c\times 2$ matrix $U$:

\beq
\left[\Phi(x-z)\right]^\alpha_i =
\frac{\rho}{\sqrt{2}\pi} e^{\mu (x_4-z_4)} \sqrt{\Pi(x-z)} \left[\dd
\left( \frac{e^{-\mu (x_4-z_4)}\Delta(x-z,\mu)}{\Pi(x-z)} \right)
\frac{1\pm\gamma_5}{2}\right]_{ij}\epsilon^{jk}U^\alpha_k\,.
\la{zeromode_x}\eeq
\beq
\Pi(x) = 1 + \frac{\rho^2}{x_4^2+r^2+\rho^2} \; ,\quad
\Delta(x,\mu) = \frac{1}{x_4^2+r^2}\left[ \cos(\mu r) + \frac{x_4}{r}
\sin(\mu r) \right]\,.
\eeq
Its Fourier transform is defined as
$\Phi(p,\mu) = i \int {\rm d}^4x\,e^{-ip\cdot x}\Phi(x,\mu)$
and has the structure
$\Phi(p,\mu) = \gamma_\alpha\varphi_\alpha(p,\mu)$.
The Lorentz symmetry is broken at finite $\mu$, and the components
become
\beqa
\varphi_4(p_4,p;\mu) &\pj=&\pj \frac{\pi\rho^2}{4p} \Big\{
(p-\mu-ip_4)\left[(2p_4+i\mu)f_{1-} + i(p-\mu-ip_4)f_{2-}\right]
\nonumber\\
&&\quad\quad+(p+\mu+ip_4)\left[(2p_4+i\mu)f_{1+} -
i(p+\mu+ip_4)f_{2+}\right]\Big\}\,,
\nonumber\\
\varphi_i(p_4,p;\mu) &\pj=&\pj \frac{\pi\rho^2 p_i}{4p^2} \Bigg\{
(2p-\mu)(p-\mu-ip_4)f_{1-}+(2p+\mu)(p+\mu+ip_4)f_{1+} \nonumber\\
&&\quad\quad\quad + \left[2(p-\mu)(p-\mu-ip_4) -
\frac{1}{p}(\mu+ip_4)[p_4^2+(p-\mu)^2] \right] f_{2-}
\nonumber\\
&&\quad\quad\quad +\left[2(p+\mu)(p+\mu+ip_4) +
\frac{1}{p}(\mu+ip_4)[p_4^2+(p+\mu)^2] \right] f_{2+}\Bigg\}\,,
\eeqa
where the scalar $p = \vert\vec p\vert$, the spatial $i=1\dots 3$,
and the functions
\beq
f_{1\pm} = \frac{I_1(z_{\pm})K_0(z_{\pm}) - I_0(z_{\pm}) K_1(z_{\pm})}{z_{\pm}}
\,, \quad
f_{2\pm} = \frac{I_1(z_{\pm})K_1(z_{\pm})}{z_{\pm}^2}
\eeq
are evaluated at $z_{\pm} = \frac{1}{2}\rho\sqrt{p_4^2+(p\pm\mu)^2}$.
With these expressions it is explicitly verified that the normalization
condition holds for any $\mu$:
\beq
1 =\int\!\frac{d^4p}{(2\pi)^4}\:\tilde\Phi_I(p,\mu)\Phi_I(p,\mu)=
\int\!\frac{d^4p}{(2\pi)^4}\:\left[ \varphi_4^*(-\mu)\varphi_4(\mu)
+ \vec\varphi^*(-\mu)\cdot\vec\varphi(\mu)\right].
\la{1}\eeq

\end{document}